\documentclass[preprint,floats,aps,epsfig,nofootinbib,amssymb]{revtex4-1}
\usepackage{mathrsfs}

\usepackage{slashed}
\usepackage{graphicx,color}
\usepackage{dcolumn}
\usepackage{bm}
\usepackage{subfig}
\usepackage{graphicx}
\usepackage{amssymb}

\def \R {\mathbf{R}}
\def \N {\mathbf{N}}

\begin{document}

\title{Phenomenology of the gauge symmetry for right-handed fermions}

\author{Wei Chao}
\email{chaowei@bnu.edu.cn}
\affiliation{Center for Advanced Quantum Studies, Department of Physics, Beijing Normal University, Beijing, 100875, China}

\vspace{3cm}

\begin{abstract}
In this paper we investigate the phenomenology of the U(1) gauge symmetry  for right-handed fermions, where three right-handed neutrinos are introduced for anomalies cancellation.
Constraints on the new gauge boson $Z_{\mathbf{R}}$ arising from $Z-Z^\prime$ mixing as well as the upper bound of $Z^\prime$ production cross section in di-lepton channel at the LHC  are presented. 
We further study the neutrino mass generation and  the phenomenology of $Z_{\mathbf{R}}$-portal dark matter in this model.
The lightest right-handed neutrino can be the cold dark matter candidate stabilized by a $Z_2$  flavor symmetry.  
Our results show that  active neutrino masses can be generated via the modified type-II seesaw mechanism; right-handed neutrino dark matter is available only for its mass at near the resonant regime of the SM Higgs and/or  the new bosons;  constraint from the dilepton search at the LHC is stronger than that from the $Z-Z^\prime$ mixing only for $g_\R<0.121$, where $g_\R$ is the new gauge coupling.

\end{abstract}

\maketitle
\section{Introduction}
Local U(1) extensions of the standard model (SM) are well-motivated new physics at the TeV-scale.  
It can be constructed either from the bottom-up approach, such as  the flavorful gauge symmetry $U(1)_{\mathbf{L}_i-\mathbf{L}_j}$~\cite{He:1991qd},  or from the top-down approach, such as $U(1)_{\mathbf{N}}$~\cite{King:2005jy} which comes from the spontaneous breaking of the $E_6$ grand unification theory. 
Many models with new U(1) gauge symmetry have been proposed addressing various problems, such as $U(1)_{\mathbf{L}_i-\mathbf{L}_j}$~\cite{He:1991qd}, $U(1)_{\mathbf{N}}$~\cite{King:2005jy},  $U(1)_{\mathbf{ B-L}}$~\cite{Mohapatra:1980qe,Marshak:1979fm,Wetterich:1981bx}, $U(1)_\mathbf{B}$~\cite{FileviezPerez:2010gw,Dulaney:2010dj}, $U(1)_{\mathbf{L}}$~\cite{FileviezPerez:2010gw,Dulaney:2010dj,Chao:2010mp}, $U(1)_{\mathbf{B+L}}$~\cite{Chao:2015nsm,Chao:2016avy} generic U(1)~\cite{Appelquist:2002mw,Ekstedt:2016wyi} etc.
For a review of various U(1) models and collider signatures of the U(1)-related gauge boson, we refer the reader to Ref.~\cite{Langacker:2008yv} for detail. 
Of various local U(1) models, the $U(1)_{\mathbf{R}}$, a gauge symmetry for right-handed fermions, is attractive for following reasons:
\begin{itemize}
\item  its anomalies can be easily cancelled by introducing three right-handed Majorana neutrinos only,
\item it may occur in left-right symmetric models~\cite{leftright} and in SO(10) models~\cite{Hewett:1988xc,Langacker:1980js}, 
\item  it may solve the vacuum stability problem~\cite{Chao:2012mx} without appealing to for extra Higgs interactions, 
\end{itemize}
 but its phenomenology was not studied in detail in any reference.

In this paper  phenomenology relevant to the $U(1)_{\mathbf{R}}$ is investigated.
We first focus on constraint on the model from $Z-Z^\prime$ mixing since there is tree-level mixing between $Z$ and $Z^\prime$  in the local $U(1)_{\mathbf{R}}$ model. 
Mixing angles of the $Z_\R$ with $Z$ and $\gamma$ as well as the mass spectrum of gauge bosons are calculated.  
It shows that the $Z-Z^\prime$ mixing puts a lower bound on $v_\Phi /v$, which is the function of  $g_{\mathbf{R}}$, where $v$ is the vacuum expectation value (VEV) of the SM Higgs, $v_\Phi$ is the VEV of the scalar singlet $\Phi$ that breaks the $U(1)_{\mathbf{R}}$ spontaneously, $g_{\mathbf{R}} $ is the gauge coupling of $U(1)_{\mathbf{R}}$, while the precisely measured $Z$ boson mass puts a  strong constraint  on $v_\Phi/v$: $v_\Phi/v>73.32$.

Then we study how to naturally realize neutrino masses in the $U(1)_{\mathbf{R}}$ model.  
The discovery of the neutrino oscillations has confirmed that neutrinos are massive and lepton flavors are mixed, which provides the first evidence for new physics beyond the SM.
Canonical seesaw mechanisms~\cite{seesawI,seesawII,seesawIII} provide a natural way in understanding the tiny but non-zero neutrino masses. 
In the $U(1)_{\mathbf{R}}$ model the mass matrix of right-handed Majorana neutrinos originates from their Yukawa couplings with $\Phi$, and is thus at the TeV-scale, such that it predicts a tiny Yukawa coupling of right-handed neutrinos with left-handed lepton doublets if active neutrino masses are generated from the type-I seesaw mechanism. 
We show that neutrino masses can be generated from the modified type-II seesaw mechanism, where the scalar triplet carries no $U(1)_{\mathbf{R}}$ charge and its coupling with other scalars breaks the $\mathbf{B-L}$ explicitly.

After that  we focus on the  phenomenology of $Z_\R$-portal dark matter. 
The fact that about $26.8\%$ of the universe is made of dark matter with relic abundance $\Omega h^2$ = 0.1189, has been firmly established, while the nature of the dark matter is still unclear.
Imposing the $Z_2$-symmetry on right-handed handed neutrinos only, the lightest right-handed neutrino $\N$ can be cold dark matter candidate.  
We study constraint on the model from the observed dark matter relic density, the exclusion limits of the spin-independent direct detection cross section mediated by $\Phi$ and $H$, as well as the spin-dependent direct detection cross section mediated by the $Z_\R$. 
It shows that right-handed neutrino dark matter is available only for its mass at near the resonance of SM Higgs, new scalar singlet and the $Z_\R$.
Finally we investigate collider signatures of the $Z_\R $ at the LHC.  
Comparing its production cross section at the LHC the with upper limits  given by the ATLAS,  we get the lower limit on the  $Z_{\mathbf{R}}$ mass, which is the function of $g_\R$. 
It  shows that the constraint from the dilepton search at the LHC is stronger than that from the $Z-Z_\R$ mixing only for $g_\R<0.121$. 

The paper is organized as follows: In section II we introduce the model in detail. 
 In section III we study constraint on the model from $Z-Z^\prime$ mixing. 
 Section IV and V are focused on the neutrino mass generation and the dark matter phenomenology, respectively. 
 We study the collider signature of $Z_{\mathbf{R}}$ in section VI. The last part is concluding remarks.

\section{The model}
\begin{table}[h]
\begin{tabular}{l|c|c|c|c|c|c|c |c}
\hline
\hline fields &$\ell  $ & $Q_L$ & $\mathbf{N}_R$ & $E_R$ &$U_R$ & $D_R$& $H$ & $\Phi$ \\
\hline$U(1)_\R$ & $0$ & $0$ & $\beta$ & $-\beta$ & $\beta$ & $-\beta$ & $\beta$  & $-2\beta$\\
\hline $Z_2$ & +&+&-&+&+&+&+&+ \\
\hline 
\hline
\end{tabular}
\caption{ Quantum numbers of various fields under the local $U(1)_{\mathbf{R}}$, $\ell_L$ and $Q_L$ are left-handed fermion doublets, $\mathbf{N}_R$ is right-handed neutrinos, $\Phi$ is the electroweak singlet scalar.  }\label{aaa}
\end{table}
We formulate our model in this section. Only right-handed fermions and the SM Higgs carry
 non-zero $U(1)_\mathbf{R}$ charge which we normalize to be multiples of ``$\beta$" . 
%
%
The SM provides the even number of fermion doublets required by the global $SU(2)_L^{}$ anomaly~\cite{globalsu2}. 
The absence of axial-vector anomalies~\cite{avector1,avector2,avector3} and the gravitational-gauge anomaly~\cite{anog1,anog2,anog3}  require that  the SM should be extended with three  right-handed neutrinos.
We list in table.~\ref{aaa} the quantum number of various fields under the $U(1)_\mathbf{R}$. The anomaly cancellations conditions are listed in Table.~\ref{anomaly}.
In the following studies, we set $\beta=1$ for simplicity. 
\begin{table}[htbp]
\centering
\begin{tabular}{l|l}
\hline anomalies~~~~~~~~~~~~~~ & anomaly free conditions~~~~~~~~~~~~~~ \\
\hline
$SU(3)_C^2 U(1)_{\mathbf{R}}^{}$:  & $- 2 (\beta) - 2 (-\beta) =0 $\\
$U(1)_Y^2 U(1)_{\mathbf{R}}^{} $: & $-\left[  3 \left( {2\over 3}\right)^2 \beta  +3 \left( {1 \over 3} \right)^2  (-\beta)+ (-1)^2 (-\beta)\right] = 0$ \\
$U(1)_{\mathbf{R}}^2 U(1)_Y$:&  $-\beta^2 \left[ 3 \times {2 \over 3} - 3 \times {1 \over 3} -1 \right]=0 $ \\
$U(1)_{\mathbf{R}}$: & $ -\left[\beta+(- \beta)\right] - 3 [ \beta+  (-\beta) ]=0 $ \\
$U(1)_{\mathbf{R}}^3$: &  $-[\beta^3 + (-\beta)^3] - 3 [\beta^3 + (-\beta)^3] =0$ \\
\hline
\end{tabular}
\caption{ The anomaly cancellation conditions of the $U(1)_{\mathbf{R}}$.  }\label{anomaly}
\end{table}

The covariant derivative  can be written as $D_\mu=\partial_\mu-ig\tau^a A_\mu^a -ig'YB_\mu -ig_\mathbf{R} Y_\R B^\prime_\mu $,  where $g_\mathbf{R}$ is the gauge coupling of $U(1)_\mathbf{R}$ and $Y_\R=0,\pm1$. 
The scalar potential and the Yukawa interactions take the following form: 
\begin{eqnarray}
V_{~}^{}&=&-\mu^2 H^\dagger H + \lambda (H^\dagger H)^2 -\mu_\Phi^2 \Phi^\dagger \Phi + \lambda_1 (\Phi^\dagger \Phi)^2 + \lambda_2 (\Phi^\dagger \Phi) (H^\dagger H) \,  \\
{\cal L}_Y &=& - \overline{Q_L^{}} Y_U^{} \widetilde{H} U_R^{} -  \overline{Q_L^{}} Y_D^{} H D_R^{} -\overline{\ell_L^{}} Y_E^{} H E_R^{} - \overline{\mathbf{N}_R^C} Y_\mathbf{N} \Phi \mathbf{N}_R^{} + {\rm h.c.} \, 
\end{eqnarray}
where  $H\equiv [G^+, (h+iG^0_h + v)/\sqrt{2}]^T$  with $v$ the VEV of $H$ and $\Phi \equiv (\phi +iG^0_s +v_\Phi)/\sqrt{2}$ with $v_\Phi$ the VEV of $\Phi$, $Y_U,~Y_D$ and $Y_E$ are $3\times3$ Yukawa matrices.
The Yukawa interaction of right-handed neutrinos with left-handed lepton doublets is forbidden by the $Z_2^{}$ discrete flavor symmetry and the lightest $\mathbf{N}_R^{}$ is the cold dark matter candidate.
Imposing the minimization conditions, one has
\begin{eqnarray}
m_{h,\phi}^2&= &  (v^2 \lambda + v_\Phi^2 \lambda_1^{} ) \mp \sqrt{ (v^2 \lambda - v_\Phi^2 \lambda_1^{} ) ^2 + v^2 v_\Phi^2 \lambda_2^2}\\
\alpha~~~ &=& {1 \over 2 } \arctan\left[  { v v_\Phi \lambda_2 \over v^2 \lambda -v_\Phi^2 \lambda_1^{} }\right]  
\end{eqnarray}
where $\alpha$ is the mixing angle between CP even scalars. 
The physical parameters in the scalar sector are $m_h$, $m_\phi$, $\alpha$, $v$ and $v_\Phi$, while all other parameters can be reconstructed by them. 
The mixing angle $\alpha$ is constrained by the data from Higgs measurements at the LHC.
Universal Higgs fits~\cite{Giardino:2013bma} to the data of ATLAS and CMS collaborations were performed in Ref.~\cite{Chao:2015nsm,Chao:2016avy,Chao:2016vfq}, and one has $\cos\alpha>0.865$ at the 95\% confidence level.  
The constraint from electroweak precision observables is usually weaker than  that from universal Higgs fit~\cite{Chao:2016cea}. 
For the beta functions of $\lambda_i$ and $g_{\mathbf{R}}$ as well as their impacts on the vacuum stability of the SM Higgs, we refer the reader to Ref.~\cite{Chao:2012mx} for detail.

\begin{figure*}[t]
\begin{center}
  \includegraphics[width=0.45\textwidth]{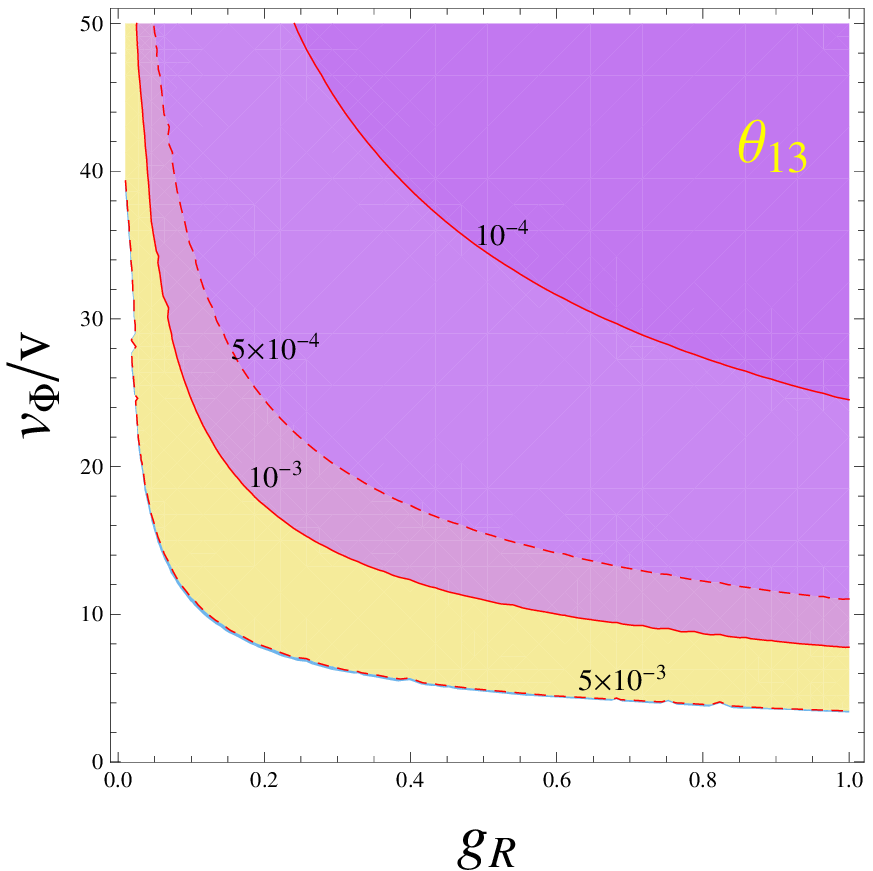} 
   \includegraphics[width=0.45\textwidth]{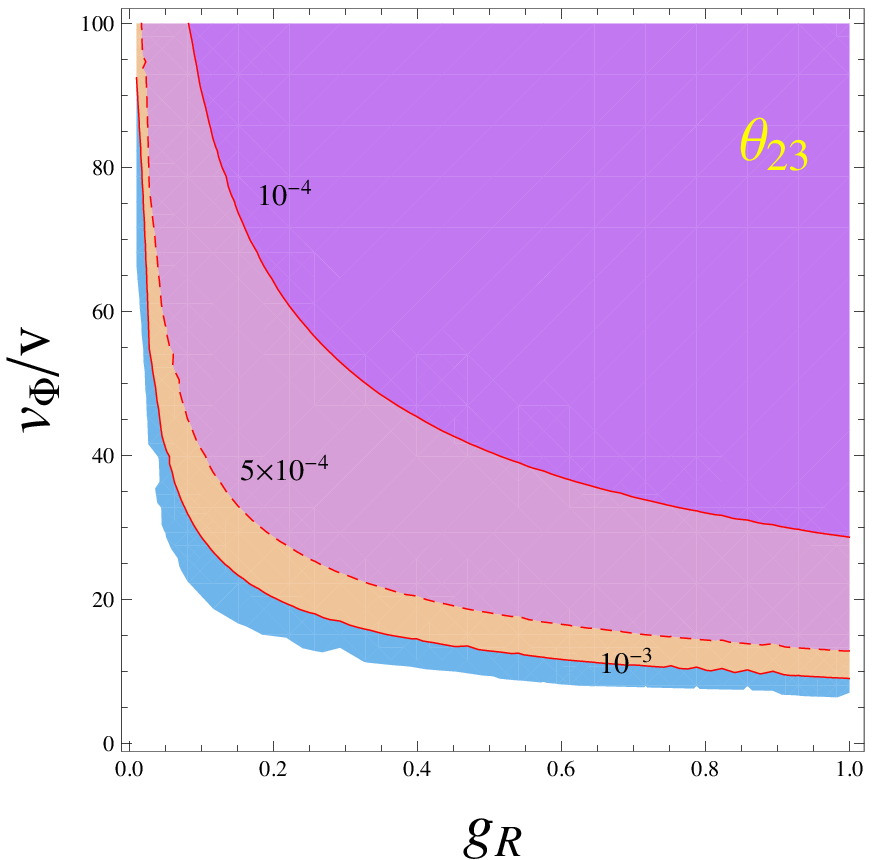} 
\caption{Contours of $\theta_{13}$ (left-panel) and $\theta_{23}$ (right-panel) in the $g_{\mathbf{R}}-v_\Phi/v$ plane.
\label{fig:theta}
}
\end{center}
\end{figure*}

\section{Vector boson Masses and Mixings}
Masses of gauge bosons come from the spontaneous breaking of the $SU(2)_L^{} \times U(1)_Y^{} \times U(1)_\mathbf{R}$ gauge symmetry. 
Since the SM Higgs carry non-zero $U(1)_\mathbf{R}$ charge, there is mixing between $Z$ and $Z_\mathbf{R}$ at the tree-level. 
The mass matrix of neutral vector bosons in the basis $(B_\mu^{},~W_\mu^3,~B_\mu^\prime)$ is given by:
\begin{eqnarray}
M_V^2 = {v^2\over 4}\left( \matrix{ g^{\prime 2} & -g^\prime g & 2 g^\prime g_{\mathbf{R}}\cr -g^\prime g & g^2 & -2 g g_{\mathbf{R}} \cr 2 g^\prime g_\mathbf{R}^{} & -2 g g_{\mathbf{R}} & 4 g_{\mathbf{R}}^2 (1 + \delta ) } \right ) \; , \label{massmatrix}
\end{eqnarray}
where  $g,~g^\prime$ are gauge couplings of $SU(2)_L$ and $U(1)_Y$ respectively;  $\delta= 4 v_\Phi^2 /v^2$.
In the limit $g_\mathbf{R}^{} \to 0$, one gets the mass matrix of of  the SM gauge bosons.
The mass matrix in Eq. (\ref{massmatrix}) can be diagonalized by the $3\times3$ unitary transformation, ${\cal U}^\dagger M_V^2 {\cal U}^* = {\rm diag} \{ 0 ,~M_Z^2,~M_{Z_{\mathbf{R}}}^2 \} $, where ${\cal U}$ can be written in terms of the standard parameterization: 
\begin{eqnarray}
{\cal U } = \left( \matrix{ c_{12} c_{13} &  s_{12} c_{13} & c_{13} \cr -c_{23} s_{12} -c_{12} s_{13} s_{23} & c_{12} c_{23} -s_{12} s_{13} s_{23} & c_{13} s_{23} \cr -c_{12} c_{23} s_{13}+s_{12} s_{23} & -c_{23} s_{12} s_{13} -c_{12} s_{23 } & c_{13} c_{23} } \right) \; ,
\end{eqnarray}
with $c_{ij} \equiv \cos \theta_{ij}$ and $s_{ij} \equiv\sin \theta_{ij}$.
Mass eigenvalues of   $Z$ and $Z_{\mathbf{R}} $ are 
\begin{eqnarray}
M_{Z_\R  (Z)}^2 = {v^2 \over 8 }\left\{ g^2 + g^{\prime 2} + 4 g_{\mathbf{R}}^2 (1+  \delta)  \pm \sqrt{-16 \delta (g^2 + g^{\prime2}) g_{\mathbf{R}}^2 +[ g^2 + g^{\prime2} + 4 g_{\mathbf{R}}^2(1+\delta)]^2} \right\} \label{eigenvalue}
\end{eqnarray}
For the case $4 g_{\mathbf{R}}^2 (1+  \delta) \gg g^2 + g^{\prime 2} $, which corresponds to the decoupled limit, it has $M_{Z^\prime}^2 \approx  v^2 (1+\delta) g_{\mathbf{R}}^2$ and $M_Z^2 \approx (g^2 + g^{\prime 2} )v^2/4$. 
Mixing angles are  actually the function of $g_{\mathbf{R}}$ and $\delta$:
\begin{eqnarray}
\tan \theta_{23}^{} & =& -{g \over 2 g_{\mathbf{R}} }\left( 1- { g_{\mathbf{R} }^2 v^2 \delta \over M_{Z_\R}^2 }\right) \; \\
\sin \theta_{13} &=& { g^\prime (M_{Z_\R}^2 -g_{\mathbf{R}}^2 v^2 \delta ) \over  \left[ (g^2 + g^{\prime 2 } ) (M_{Z_\R }^2 - g_{\mathbf{R}}^2 v^2 \delta )^2 + 4 g_{\mathbf{R}}^2 M_{Z_\R }^4\right]^{1/2} } \\
\tan \theta_{12} &= & { g^\prime (M_{Z}^2 -g_{\mathbf{R}}^2 v^2 \delta ) \sqrt{g^2 +g^{\prime 2}} \over g  \left [ (g^2 + g^{\prime 2 } ) (M_{Z }^2 - g_{\mathbf{R}}^2 v^2 \delta )^2 + 4 g_{\mathbf{R}}^2 M_{Z }^4 \right ]^{1/2} }
\end{eqnarray}
where $\theta_{12}^{} $ corresponds to the conventional weak mixing angle, $\theta_{13}$ and $\theta_{23}$ are mixing angles of $\gamma-Z_\mathbf{R}^{}$ and $Z-Z_\mathbf{R}^{}$ respectively.

\begin{figure*}[t]
\begin{center}
  \includegraphics[width=0.5\textwidth]{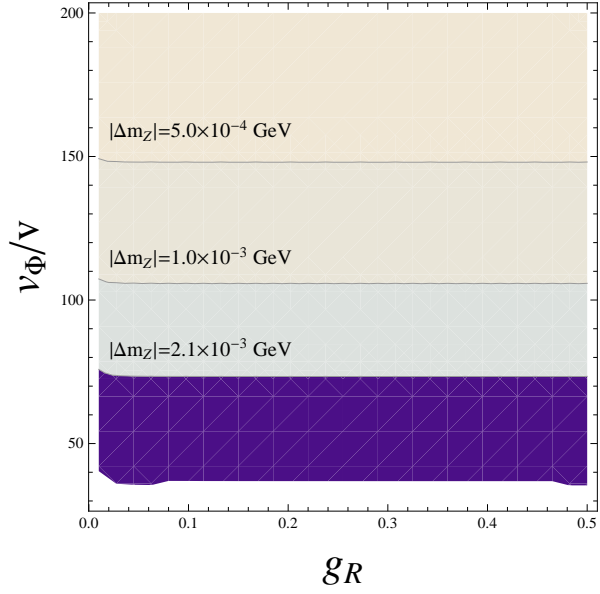}
\caption{Contours of $\Delta M_Z$ in the $g_\mathbf{R}-v_\Phi/v$ plane.
\label{fig:zmass}
}
\end{center}
\end{figure*}

Due to  the $Z-Z_\mathbf{R}$ mixing, $Z_\mathbf{R}$ may decay into charged gauge boson pairs $W^-W^+$, which process can be used to place constraint on the $Z-Z_\mathbf{R}$ mixing using diboson production at the LHC. 
It shows that the $Z-Z_\mathbf{R}$ mixing should be less than $0.7\sim2\times 10^{-3}$~\cite{Osland:2017ema} derived from the data recoded by ATLAS and CMS  collaborations at $\sqrt{s}=13~\text{TeV}$ with integrated luminosities of $13.2~{\rm fb^{-1}}$ and $35.9~{\rm fb}^{-1}$, respectively. 
%
%
In Fig.~\ref{fig:theta}, we show  contours of the $\theta_{13}$ (left-panel) and $\theta_{23}^{}$ (right-panel) in the $g_{\mathbf{R}}-v_\Phi/v$ plane. 
We take $\alpha(M_Z)^{-1}=127.918$, $\sin^2 \theta_W (M_Z^{})=0.23122$ and $M_Z=91.1876~{\rm GeV}$~\cite{pdg}, which are used to get values of $g$ and $g^\prime$ respectively.
One can see from the right-panel of Fig.~\ref{fig:theta} that, the scale  of the $U(1)_{\mathbf{R}}$ breaking should be at least one order higher than the electroweak scale.
According to Eq. (\ref{eigenvalue}), the Z-boson mass is slightly changed in the $U(1)_\mathbf{R}^{}$ model. 
There is thus constraint from the precision measurement of the $Z$ boson mass.
We show in the Fig.~\ref{fig:zmass} contours of $\Delta M_Z^{} $, namely $M_Z^{}-M_Z^{\rm observed}$, in the $g_\mathbf{R}^{} -v_\Phi/v$ plane. 
It shows that $\Delta M_Z^{} $ is insensitive to the $g_{\mathbf{R}}$ when it is larger than $0.04$.
Using the ambiguity of the Z-boson mass given by the PDG~\cite{pdg}, which is $\Delta M_Z^{}<0.0021~\text{GeV}$, one has $v_\Phi/v>73.32$ ($v_\Phi >18~\text{TeV}$).

\section{Neutrino masses}
In this section we investigate how to generate Majorana masses of active neutrinos in the $U(1)_\mathbf{R}$ model. 
The solar, atmosphere, accelerator and reactor neutrino oscillation experiments have firmed that neutrinos are massive and lepton flavors are mixed. 
In our model right-handed neutrinos  do not couple to left-handed lepton doublets, so that the conventional type-I seesaw mechanism does not work. 
We study the possibility of generating active neutrino masses via the type-II seesaw mechanism. 
Interactions relevant to the scalar triplet $\Delta$ with $Y=2$ can be written as
\begin{eqnarray}
-{\cal L}_\Delta^{} = M_\Delta^2 \Delta^\dagger \Delta +\left(  \tilde \lambda  H^T i\sigma_2^{} \Delta H \Phi +{\rm h.c. }\right)+ \left( \overline{\ell_L^{} } Y_\Delta^{} \Delta \ell_L^{} +{\rm h.c.} \right)+\cdots \label{neutrino}
\end{eqnarray}
where dots stand for interactions of $\Delta$ that are irrelevant to the neutrino mass generation, $Y_\Delta$ is the symmetric Yukawa coupling matrix. 
Full expression of triplet interactions can be found in Ref.~\cite{Chao:2006ye}.  
After the spontaneous breaking of the $SU(2)_L^{} \times U(1)_Y^{} \times U(1)_\mathbf{R}$ symmetry, the active neutrino masses can be written as
\begin{eqnarray}
M_\nu = Y_\Delta^{} v_\Delta^{} \approx Y_\Delta \tilde \lambda {v^2 v_\Phi \over M_\Delta^2 }
\end{eqnarray}
where $v_\Delta$ is the VEV of $\Delta$.  
The smallness of active neutrino masses is own to the tiny $v_\Delta$, which is suppressed by the $M_\Delta^2$.  
Notice that interactions in the first bracket of Eq.~(\ref{neutrino}) breaks the lepton number $\mathbf{L}$ explicitly and the global $U(1)_\mathbf{L}$ can be recovered in the limit $\tilde \lambda \to 0$.
According to the naturalness criterion~\cite{thooft}, $\tilde \lambda$ should be naturally small.  
In this case $\Delta$ can be at the TeV scale and we refer the reader to Refs.~\cite{Perez:2008ha,Chao:2008mq,Chao:2009ef} for signatures of $\Delta$ at the LHC.

\section{Dark matter}
About $26.8\%$ of our universe is made of dark matter, whose relic abundance is measured as $\Omega_{\rm DM} h^2 =0.1198\pm0.0015$~\cite{Aghanim:2015xee}. 
Weakly Interacting Massive Particle (WIMP)~\cite{Arcadi:2017kky} is a promising dark matter candidate, since the observed relic density can be naturally derived for a WIMP mass around the electroweak scale.
In the $U(1)_\mathbf{R}$ model, the lightest right-handed Majorana neutrino can be cold dark matter candidate, stabilized by the $Z_2$ symmetry\footnote{For the vector-boson portal neutrino dark matter, we refer the reader to Refs.~\cite{Okada:2010wd,Chao:2012sz,Chao:2012pt,Okada:2016gsh,Okada:2016tci} for detail.}. 
We evaluate the relic abundance of the dark matter and study its implications in dark matter direct detections in this section.
The dark matter  $\mathbf{N}$ mainly couple to the $Z_\mathbf{R}$, $\hat s$ and $\hat h$, with Lagrangian: 
\begin{eqnarray}
-{\cal L} \sim {1\over 2  }\overline \mathbf{N} ig_{\mathbf{R}}^{}  \gamma_\mu^{} \gamma_5^{} Z_{\mathbf{R}}^\mu \mathbf{N} + {1\over 2 } {m_{\mathbf{N}} \over v_\Phi} \overline{ \mathbf{N}} (c_\alpha \hat s- s_\alpha \hat h) \mathbf{N}
\end{eqnarray}
where $\mathbf{N} =N_R^C + N_R^{}$, $m_{\mathbf{N}}$ is the mass eigenvalue of $\mathbf{N}$, $\hat h$ and $\hat s$ are the mass eigenstates of $h$ and $s$ respectively. 
The interaction of $\mathbf{N}$ with $Z$ can be neglected due to the tiny mixing angle $\theta_{23}$.

\begin{figure*}[t]
\begin{center}
  \includegraphics[width=0.6\textwidth]{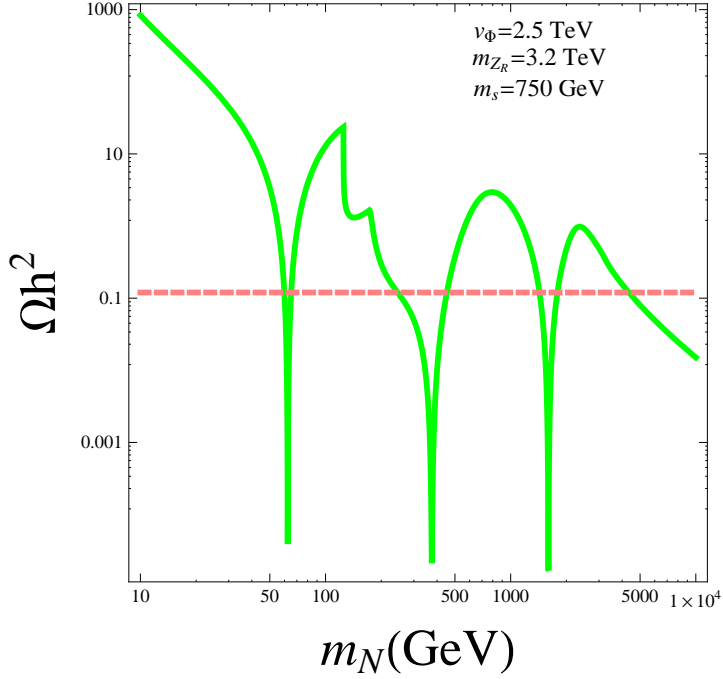} 
\caption{The relic density of the dark matter as the function of the dark matter mass $m_\mathbf{N}$, by setting $c_\alpha=0.9$, $m_s=750~{\rm GeV}$, $v_\Phi =2.5~{\rm TeV}$ and $m_{Z_\R}=3.2~{\rm TeV}$. The horizontal line is the observed dark matter relic density.
\label{relic}
}
\end{center}
\end{figure*}

The evolution of the dark matter density $n$ is governed by the thermal average of reduced annihilation cross sections $\langle \sigma v\rangle$, which can be approximated with the non-relativistic expansion: $\langle \sigma v\rangle = a + b \langle v^2\rangle$. 
Contributions of various channels can be written as
\begin{eqnarray}
&\langle \sigma v \rangle_{s_a s_b} =&  {1\over 1 + \delta } {\lambda^{1/2} (4, \boldsymbol{\zeta}_a^{}, \boldsymbol{\zeta}_b^{}) \over 1024 \pi v_\Phi^2 m_\mathbf{N}^2 } \left| {c_\alpha^{}  C_{s ab} \over 4-\boldsymbol{\zeta}_s^{}} -  { s_\alpha^{} C_{hab}\over 4 -\boldsymbol{\zeta}_h^{}} \right|^2 \langle v^2 \rangle   \\
%
%
%
%
&\langle \sigma v \rangle_{ZZ} = &
\frac{s_{2\alpha}^2 m_Z^4}{512\pi m_\mathbf{N}^2 v^2 v_\Phi^2} \sqrt{1-\boldsymbol{\zeta}_Z^{}} \left ( 3-{4 \over \boldsymbol{\zeta}_Z^{}} + {4  \over \boldsymbol{\zeta}_Z^{2}}\right)
\left| { 1 \over 4-\boldsymbol{\zeta}_s^{}} -  { 1 \over 4 -\boldsymbol{\zeta}_h^{}} \right|^2
\langle v^2  \rangle,  \\
%
%
%
&\langle \sigma v \rangle_{WW} = & \frac{s_{2\alpha}^2 m_W^4}{256\pi m_\mathbf{N}^2 v^2 v_\Phi^2 } \sqrt{1-\boldsymbol{\zeta}_W^{}} \left ( 3-{4 \over \boldsymbol{\zeta}_W^{}} + { 4 \over \boldsymbol{\zeta}_W^{2}} \right)
\left| { 1 \over 4 -\boldsymbol{\zeta}_s^{}} -  { 1\over 4 -\boldsymbol{\zeta}_h^{}} \right|^2
\langle v^2  \rangle    \\
%
%
%
%
%
%
%
%
&\langle \sigma v\rangle_{f\bar f} = &   \frac{n_C^f s_{2\alpha}^2 \boldsymbol{\zeta}_f^{}}{32 \pi   v^2 }(1- \boldsymbol{\zeta}_f^{})^{3/2}
 \left| { 1 \over 4 - \boldsymbol{\zeta}_s^{}} - { 1 \over 4 -\boldsymbol{\zeta}_h^{}} \right|^2 \langle v^2 \rangle + \nonumber \\
 && { n_C^f g_{\mathbf{R}}^4\over 192 \pi m_\mathbf{N}^2 } \sqrt{1-\boldsymbol{\zeta}_f^{}} \left(\boldsymbol{\zeta}_f^{} + 2\right)      \left|{1  \over 4 -\boldsymbol{\zeta}_{Z_\mathbf{R}}^{}}\right|^2 \langle v^2 \rangle + { n_C^f g_{\mathbf{R}}^4  \varrho_{Z_\mathbf{R}}^f \sqrt{1 -\boldsymbol{\zeta}_f^{}} \over 32 \pi m_{Z_\mathbf{R}}^2} + \nonumber \\
 && {   23\boldsymbol{\zeta}_f^{2}   -192 \varrho_{Z_\mathbf{R}}^f  \boldsymbol{\zeta}_{Z_\mathbf{R}}^{-1}  + 8(30 \varrho_{Z_\mathbf{R}}^{f2} + 12 \varrho_{Z_\mathbf{R}}^f+1 ) - 4 \boldsymbol{\zeta}_f^{} (30 \varrho_{Z_\R}^f+7) \over  768 \pi m_\mathbf{N}^2 \sqrt{1-\boldsymbol{\zeta}_f^{} }}  { n_C^f g_{\mathbf{R}}^4 \langle v^2 \rangle  \over \left|  4-\boldsymbol{\zeta}_{Z_\mathbf{R}}^{}  \right|^2 }  \\
 &\langle \sigma v\rangle_{Z's}^{}  =&{ c_\alpha^2 g_{\mathbf{R}}^2 m_{Z^\prime}^4\lambda^{3/2} (4, \boldsymbol{\zeta}_V^{}, \boldsymbol{\zeta}_s^{}  ) \over  1024 \pi  v_\Phi^2 m_\mathbf{N}^4 \boldsymbol{\zeta}_{Z_\mathbf{R}}^{3} }+ {\cal O} (\langle v^2\rangle) \label{v2vs}
\end{eqnarray}
where $\lambda(x,~y,~z)\equiv x^2+y^2+z^2-2xy-2xz-2yz$, $\boldsymbol{\zeta}_X^{} =m_X^2 /m_\mathbf{N}^2 $, $\varrho_{Z_\R}^f = m_f^2 /m_{Z_\R}^2$;  $\delta_{ab}^{}=1$ (for $a=b$) and $0$ (for $a\neq b$);   $C_{s_i s_j s_k}$ are trilinear couplings; $s_{2\alpha} =\sin 2 \alpha$, $c_\alpha=\cos \alpha$, $n_C^f$ is the color factor of $f$.

The final relic density can be given by~\cite{Bertone:2004pz}
\begin{eqnarray}
\Omega_{\rm DM} h^2 = {1.07\times 10^9 \over M_{pl} } {x_F \over \sqrt{g_\star}} {1 \over a + 3 b/x_F}
\end{eqnarray}
where $M_{pl} =1.22\times 10^{19}~\text{GeV} $ being the Planck mass, $x_F = m_\mathbf{N}/T_F$ with $T_F$ the freeze-out temperature, $g_\star$ is the effective degree of freedom at the freeze-out temperature.
As an illustration, we show in Fig.~\ref{relic} the dark matter relic density as the function of the dark matter mass $m_\mathbf{N}$ by setting  $c_\alpha=0.9$, $m_s=750~{\rm GeV}$, $v_\Phi =2.5~{\rm TeV}$ and $m_{Z_\R}=3.2~{\rm TeV}$.
The value of $g_\mathbf{R}$ is determined by $g_\mathbf{R} \approx M_{Z_\mathbf{R}}/\sqrt{v^2 +4 v_\R^2}$.
The horizontal line is the experimental value of the relic density.
One can conclude from the plot that $\mathbf{N}$ is qualified dark matter only when its mass lies near the resonance of $\hat h$, $\hat s$ or $Z_\mathbf{R}$.

\begin{figure*}[t]
\begin{center}
  \includegraphics[width=0.45\textwidth]{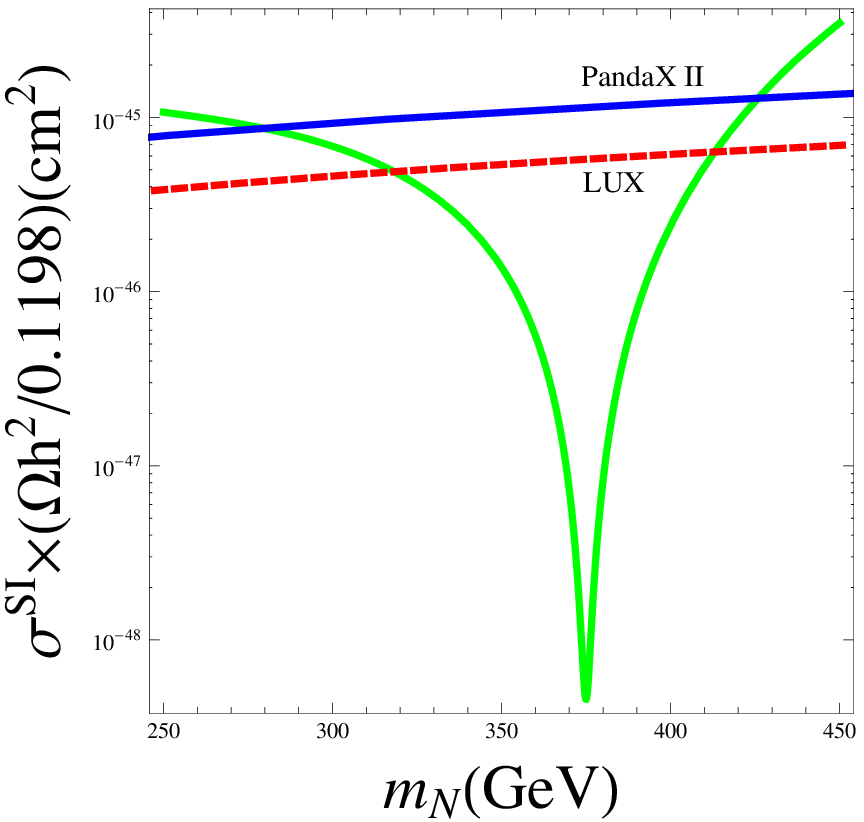} 
   \includegraphics[width=0.48\textwidth]{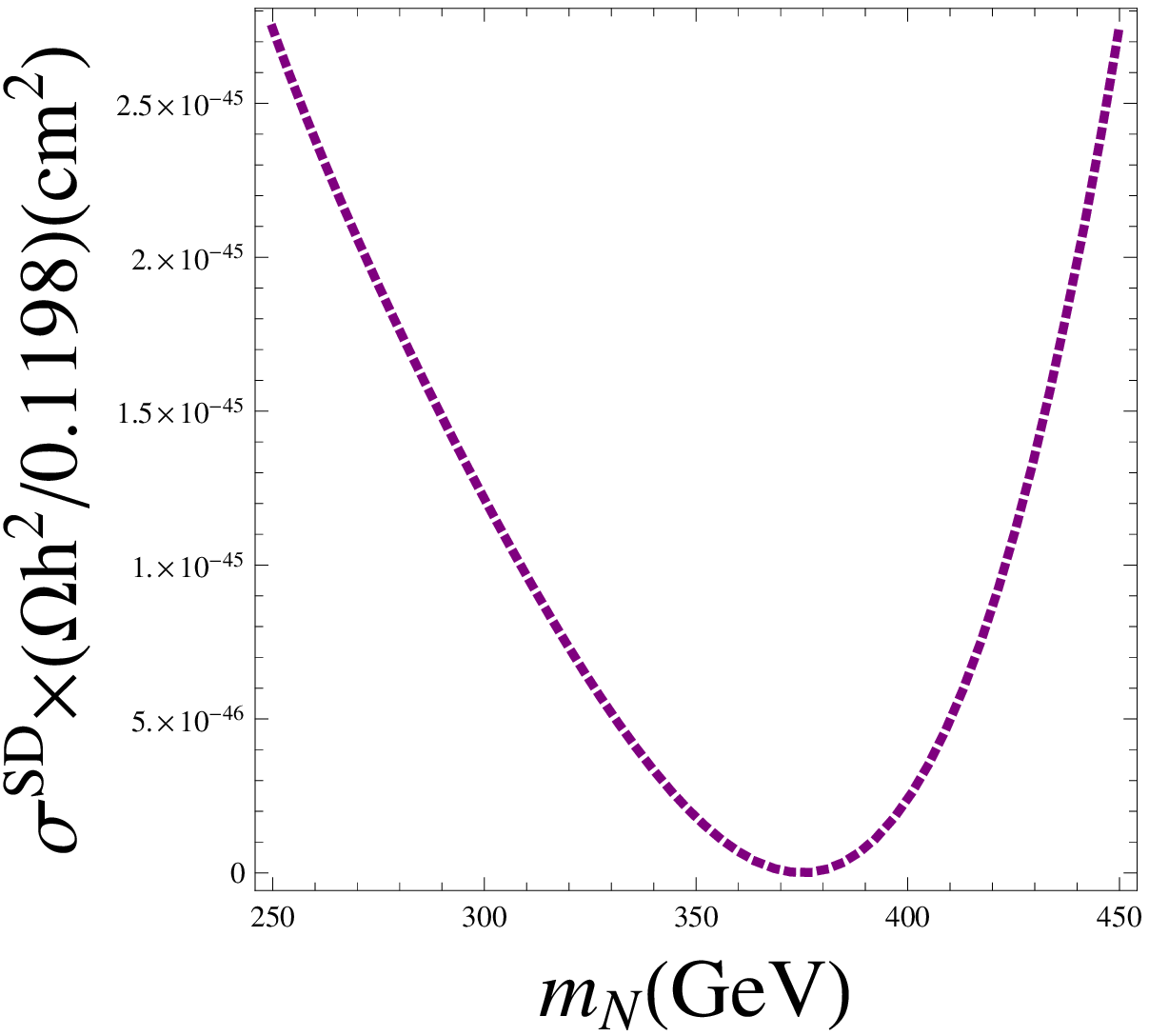} 
\caption{The rescaled spin-independent cross section (left-panel) and spin-dependent cross section as the function of the dark matter mass, other inputs are the same as these in the Fig.~\ref{relic}.
\label{fig:DD}
}
\end{center}
\end{figure*}

The spin-independent $\N-$nucleus scattering cross section is 
\begin{eqnarray}
\sigma_{\rm SI} = {\mu^2 s_{2\alpha}^2  m_N^2 \over 4 \pi v^2 v_\Phi^2} \left( {1\over m_h^2} -{1\over m_s^2}\right)^2 [Z f_p^{} + (A-Z) f_n^{} ]^2
\end{eqnarray}
where $\mu$ is the reduced mass of $\N$-nucleus system, $f_{p,n}^{}=m_{p,n} (2/9+7/9\sum_{q=u,d,s} f_{T_q}^{p,n})$ with $f_{T_u}^p=0.020\pm 0.004$, $f_{T_d}^p=0.026\pm 0.005$, $f_{T_u}^n=0.014\pm 0.003$, $f_{T_d}^n=0.036\pm 0.008$ and $f_{T_s}^{p,n}=0.118\pm0.062$~\cite{Ellis:2000ds}.
We show in the left panel of Fig.~\ref{fig:DD} the rescaled spin-independent cross section, namely $\sigma_{\rm SI} \times \Omega_{\rm DM} h^2 /0.1189$, as the function of $m_\N$ with the same inputs of the Fig.~\ref{relic}. 
We only focus on the resonant regime of $\hat s$ in this plot since there is redundancy relic density in other mass region and the resonant regime of $Z_\R$ ($\hat h$) predicts a too heavy (light) dark matter.
The red-dashed and blue-solid lines are the exclusion limits of LUX 2016~\cite{Akerib:2016vxi} and PandaX II~\cite{Tan:2016zwf} respectively.  
The available parameter space is shrunk compared with the relic density allowed region.

Since $\N$ couples to $Z_\R$, there is spin-dependent cross section of $\N$ with nucleus which takes the following form
\begin{eqnarray}
\sigma^{\rm SD} = {g_{\mathbf{R}}^4 \mu^2 \over \pi M_{Z_\R}^4 }  \left( \sum_{q=u,d,s} \lambda_q^{} \right)^2 J_N^{} (J_N^{}+1) \; ,
\end{eqnarray}
where $\lambda_q$ reduces to $\Delta^{p,n}$  for scattering off free proton or neutron with $\Delta^p_u=0.78\pm0.02$, $\Delta^p_d=-0.48\pm0.02$, $\Delta^p_s=-0.15\pm0.02$, $\Delta^n_u=-0.48\pm0.02$, $\Delta^n_d=-0.78\pm0.02$, $\Delta^n_s=-0.15\pm0.02$~\cite{Ellis:2000ds,Mallot:1999qb,Agrawal:2010fh}, $J_N$ is the angular momentum of the nucleus and it equals to $1/2$ for free nucleons. 
We show in the right-panel of the Fig.~\ref{fig:DD} the rescaled spin-dependent cross section of $\N$ with free neutron as the function of $m_\N$.  
The smallest excluded WIMP-neutron cross section is $\sigma^{\rm SD}_n=4.3\times 10^{-41}$ at $m_{\rm DM}=45~\text{GeV}$ from the PandaX-II~\cite{Fu:2016ega}. 
One can conclude from the plot that the spin-dependent cross section at the resonant regime of $\hat s$ is much smaller than the current exclusion limit.

\begin{figure*}[t]
\begin{center}
  \includegraphics[width=0.5\textwidth]{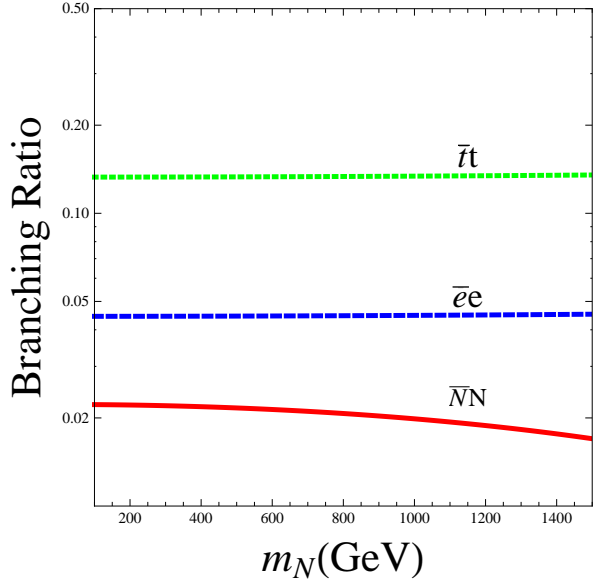} 
\caption{Branching ratios of $Z_\R$ decaying into various final states.
\label{branching}
}
\end{center}
\end{figure*}

\section{Collider signatures}
In this section we investigate the  signature of $Z_\R$ at the LHC.
$Z_\R$ can be produced at the LHC via the Drell-Yan process, while the diphoton channel provides a significant signature. 
We first study the branching ratio of $Z_\R$. 
The decay rate of $Z_\R $ into fermion pairs can be written as
\begin{eqnarray}
\Gamma(\mathbf{Z}_\R \to \bar f f) = {1\over 1+\delta_f} {n^C_f g_{\mathbf{R}}^2 m_{\mathbf{Z}_\R}^{}  \over 24 \pi } \left( 1- {m_f^2 \over m_{\mathbf{Z}_\R}^2 } \right)^{3/2} \; ,
\end{eqnarray}
where  $\delta_f=1(0)$ if $f$ is (not) identical particle. 
$Z_\R$ can also decay into diboson pair ($W^+W^-$) due to its mixing with $Z$, which is suppressed by $\theta_{23}^2$ and will be neglected in the following analysis.
We show in the Fig.~\ref{branching} the branching ratio of $Z_\R$ decaying into various final states as the function of $m_\N$ by setting $m_{Z_\R}=3.6~\text{TeV}$,  where the solid, dashed and dotted lines correspond to the decay channel of $\overline\N\N$, $\bar e e$ and $\bar t t$ respectively. 
It shows that the branching ratio of dilepton with discrete flavor is about $4.5\%$.

\begin{figure*}[t]
\begin{center}
 \includegraphics[width=0.5\textwidth]{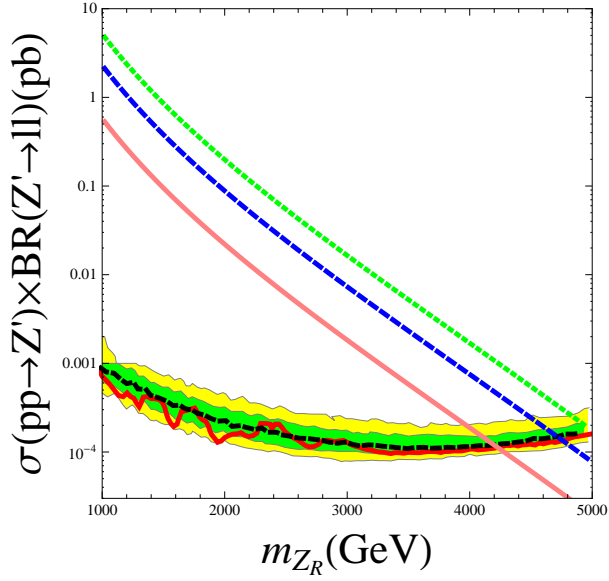} 
\caption{Production cross section of process ($pp\to Z_\R \to \ell^+ \ell^-$) as the function of $M_{Z_\R}$, the solid, dashed and dotted lines correspond to $g_\R=0.1,~0.2$ and $0.3$ respectively.
\label{zpcollider}
}
\end{center}
\end{figure*}

%
The cross section of the process $pp\to Z_\R \to \ell^+ \ell^-$ can be written as 
\begin{eqnarray}
\sigma(pp\to Z_\R \to \ell^+ \ell^-) = {3 \over s} {\Gamma_{\rm tot} \over M_{Z_\R}} \sum_q C_{q\bar q } {\rm BR} (Z_R \to q \bar q ) \times {\rm BR} (Z_\R \to \ell^+\ell^-)
\end{eqnarray}
where $\sqrt{s}$ is the centre-of-mass energy, $\Gamma_{\rm tot}$ is the total width of $Z_\R$,  $ {\rm BR} (Z_R \to q \bar q ) $ and ${\rm BR} (Z_\R \to \ell^+\ell^-)$ are branching ratios of $Z_\R$ decaying to $q\bar q $ and $\ell^+ \ell^-$ respectively,  $C_{q\bar q}$ are dimensionless partonic integrals with~\cite{Barger:1987nn}
\begin{eqnarray}
C_{q\bar q} = {4 \pi^2 \over 9} \int_{M^2/s}^1 {dx \over x } \left\{    q (x) \bar{q} \left( {M^2 \over s x}\right) + \bar q (x)  q \left( {M^2 \over s x}\right)\right\} \; . \nonumber 
\end{eqnarray}
Their numerical values are calculated with the NNPDF~\cite{Ball:2012cx} at $M=m_{Z_\R}$. 
We show in Fig.~\ref{zpcollider} the $\sigma(pp\to Z_\R \to \ell^+ \ell^-)$ as the function of $m_{Z_\R}$ where the solid, dashed and dotted lines correspond to $g_\R=0.1,~0.2$ and $0.3$ respectively. 
The green and yellow bands are upper limits on the $\sigma(pp\to Z_\R \to \ell^+ \ell^-)$ at the $1\sigma$ and $2\sigma$ separately, given by the ATLAS collaboration~\cite{Aaboud:2017buh} with $36.1~{\rm fb}^{-1}$ of proton-proton collision data collected at $\sqrt{s}=13~\text{TeV}$.
Since the $v_\Phi$ is constrained to be $v_\Phi >18~\text{TeV}$, there is lower bound on the $Z_\R$ mass from the $Z-Z_\R$ mixing.
By matching these two constraints, we find that the constraint from the dilepton search at the CERN LHC is stronger than that from the $Z-Z_\R$ mixing derived from diboson search at the LHC and the precision measurement of $Z$ boson mass only for $g_\R<0.121$ ($M_{Z_R}<4.37~\text{TeV}$).

\section{Conclusion}
Although the $U(1)_{\mathbf{R}}$ extension to the SM shares the same merit as the $U(1)_{\mathbf{B-L}}$ extension of the SM on anomalies cancellation, its phenomenology was not investigated in detail in any reference except its effect in the vacuum stability of the SM Higgs.  
In this paper we constrained the parameter space of the model using the updated results of the $Z-Z^\prime$ mixing as well as  the search of new resonance in dilepton channel at  the LHC.  
Our investigation shows that the constraint from the dilepton search at the LHC is stronger than that from the $Z-Z_\R$ mixing derived from diboson search at the LHC and the precision measurement of $Z$ boson mass only for $g_\R <0.121$.
We further studied the phenomenology of the $Z_\R$-portal dark matter and the possibility of generating active neutrino masses in the same model. 
It shows that the right-handed neutrino dark matter is self-consistent only for its mass at near the resonant regime of $\hat h$, $\hat s$ and $Z_\R$; the Majorana masses of active neutrinos can be generated from the modified type-II seesaw mechanism.
It should be mentioned that the collider signature of $\hat s$ is  Di-Higgs in various channels.
We refer the reader to Ref.~\cite{Huang:2017jws} for the Di-Higgs searches at various colliders for detail.
$\hat s$ may also be searched at the LHC via the $pp\to \hat s\to Z_\R Z_\R \to \ell_\alpha^+ \ell_\alpha^- \ell_\beta^+\ell^-_\beta$ process if $m_{\hat s} > 2 m_{Z_\R}$.

 \noindent {\bf [Note added]:} 
 When this paper was being finalized, the paper \cite{Nomura:2017tih} appeared, which partially overlaps with this one in discussing constraint from the precision measurement of Z-boson mass, They use approximated formulae when do this analysis, while we present both full analytical and numerical results in this paper.  
Their study is largely complementary to ours.

\begin{acknowledgments}
We thank to Huai-ke Guo for his help on numerical calculations and to Xiaohui Liu for helpful conversation. 
\end{acknowledgments}

\end{document}